\documentclass [
 reprint,
amssymb,
 aps,
]{revtex4-2}

\usepackage{graphicx}% Include figure files
\usepackage{dcolumn}% Align table columns on decimal point
\usepackage{bm}
\usepackage{amsmath}
\usepackage{upgreek}
\usepackage{booktabs}% bold math
\pdfoutput=1
\usepackage[mathlines]{lineno} % Enable numbering of text and display math
\begin{document}

\preprint{APS/123-QED}

\title{Tunable valley and spin splittings in $\rm \bold {VSi_2N_4}$ bilayer}% Force line breaks with \\

\author{ Li Liang}
\author{ Ying Yang}
\author{ Xiaohui Wang}
\author{Xiao Li}
\thanks{lixiao@njnu.edu.cn}
%\author{，Ying Yang, Xiaohui Wang， Xiao Li}%
\affiliation{%
Center for Quantum Transport and Thermal Energy Science, School of Physics and Technology, Nanjing Normal University, Nanjing 210023, China
}%


%author{Charlie Author}
%\affiliation{
%\affiliation{
 %Third institution, the second for Charlie Author
%\affiliation{%


\date{\today}% It is always \today, today,

\begin{abstract}
The control and manipulation of the valley and spin degrees of freedom have received great interests in fundamental studies and advanced information technologies. Compared with magnetic means, it is highly desirable to realize more energy-efficient electric control of valley and spin. Using the first-principles calculations, we demonstrate tunable valley and spin degeneracy splittings in $\rm VSi_2N_4$ bilayers, with the aid of the layered structure and associated electric control. Depending on different interlayer magnetic couplings and stacking orders, the $\rm VSi_2N_4$ bilayers exhibit a variety of combinations of valley and spin degeneracies. Under the action of a vertical electric field, the degeneracy splittings become highly tunable for both the sign and magnitude. As a result, a series of anomalous Hall currents can be selectively realized with varied indices of valley and spin. These intriguing features offer a practical way for designing energy-efficient devices based on the couplings between multiple electronic degrees of freedom.

%\begin{description}
%\end{description}
\end{abstract}

\maketitle

%\tableofcontents
\section{\label{sec:level1}Introduction}
Valley degree of freedom and associated manipulations have become rising topics in recent years \cite{2016Valley,2007Valley,2012Valley}. As local extrema in band structures, multiple valleys are selectively expressed through their contrasting optoelectronic properties. Besides introducing external magnetic field or magnetic proximity effect \cite{MF1,MF2,MF3,MF4,PRB15}, a vertical electric field, as more energy-efficient means, has been applied in a number of two-dimensional valley materials to modify the valley structure \cite{bilayerMoS2,twisted-WSe2,TiSiCO}. However, the study on the electric control of magnetic valley materials has few examples, and the influences of magnetic order and the stacking configuration remain unclear, which are worth further investigation.

A new family of two-dimensional materials, $\rm MA_2Z_4$ (M=transition metal; A=Si, Ge; Z=N, P, As), has recently proposed and synthesized \cite{MA2Z4,2MA2Z4}. As a typical example, the 
$\rm VSi_2N_4$ monolayer is a ferromagnetic, two-valley semiconductor, in which the valley splittings appear due to the simultaneous presence of the magnetic order and sizable spin-orbit coupling \cite{VSi2N4Monolayer,VSi2N4}. Further considering the $\rm VSi_2N_4$ bilayer, it will provide a powerful platform for studying the interplay of spin, valley and layer degrees of freedom.

In this work, we investigate magnetic and electronic properties of $\rm VSi_2N_4$ bilayers by first-principles density functional theory calculations. Taking into account different interlayer magnetic couplings and stacking orders, the $\rm VSi_2N_4$ bilayers exhibit varied valley and spin degeneracies. In the most stable $\rm AA^{\prime}$ stacking, the bands are spin-degenerate with sizable valley splitting in the electronic structure of the interlayer antiferromagnetic coupling, while the valley degeneracy is kept in the single-spin band structures for the ferromagnetic coupling. The introduction of a vertical electric field leads to highly tunable splittings of the above degeneracies, owing to the spin-layer and valley-layer couplings. Meanwhile, the modifications of the electronic structures give rise to distinct transport behaviors, i.e. there is a transition from the valley/spin contrasting Hall currents to single anomalous Hall current with any combination of valley and spin. The valley-spin physics from the other stacking orders is further discussed. The above characteristics are also expected in more bilayers and multilayers of magnetic valley semiconductors. The tunable valley and spin splittings from the layer degree of freedom, including the interlayer magnetic orders and the stacking orders, add a new dimension to the valley-spin physics and provide a practical avenue for designing advanced spintronic and valleytronic devices.

\section{METHODS}
We use density functional theory calculations within the generalized gradient approximation (GGA) to study the atomic and electronic structures of bilayer $\rm VSi_2N_4$ \cite{density,GGA}.The calculations are implemented in the Vienna Ab initio Simulation Package, with the projector-augmented wave potentials and the Perdew–Burke–Ernzerhof exchange-correlation functional \cite{VASP1,VASP2,PBE}. A plane-wave cutoff of 500 eV and a Monkhorst–Pack $ \bold k $-point mesh of 15 $\times$ 15 $\times$ 1 are adopted. A vacuum slab of approximately 20 $ \rm \AA $ is inserted to minimize the interaction between the $\rm VSi_2N_4$ bilayer and its periodic images.
Structure optimization is performed with a convergence threshold of $10^{-4}$ eV/$ \rm \AA $ on the interatomic forces, and the convergence criterion of the total energy for the electronic iteration is set to $10^{-8}$ eV. 
To better describe the on-site electron-electron interaction, the GGA+U method is applied to 3$d$ orbitals of the vanadium atom, with an effective $U$ of 3 eV \cite{GGAU,VS2bilayer,VSi2N4Monolayer}. The van der Waals correction is added by the DFT-D2 method to better describe the interlayer interaction of the $\rm VSi_2N_4$ bilayer \cite{DFTD2}. The spin-orbit coupling (SOC) is further introduced into the electronic structure calculations, where the out-of-plane magnetization is considered to realize the valley polarization in each $\rm VSi_2N_4$ monolayer \cite{VSi2N4Monolayer}.

To quantify the valley and spin degeneracy splittings in electronic band structures of the $\rm VSi_2N_4$ bilayers, we define the valley splitting as
\begin{eqnarray}
\Delta_{\text {val }}^{v / c}=E_{\text {val }}^{v / c, K+}-E_{\text {val }}^{v / c, K-}
\end{eqnarray}
%\end{equations}
for a given band (the valence or conduction band), and the spin splitting as 
%\begin{equations}
\begin{eqnarray}
\Delta_{\text {spin }}^{v / c, K_{\tau}}=E_{\uparrow}^{v / c, K_{\tau}}-E_{\downarrow}^{v / c, K_{\tau}} 
\end{eqnarray}
%\end{equations}
for a given band at a given valley.
Here, the superscripts, $v$ and $c$, denote the valence and conduction bands, respectively. The index, $\tau=\pm 1$, distinguishes $K_{\pm}$ valleys. ↑ and ↓ represent the spin-up and spin-down states, respectively. 
$E_{\text {val }}^{v / c, K \tau}$ in Eq. (1) is the energy extremum of the band-edge states at $K_{\tau}$ valley for a certain band regardless of the spin, while $E_{\uparrow / \downarrow}^{v / c, K\tau}$ in Eq. (2) is the energy of the band-edge state with up/down spin. 

\section{RESULTS}

\subsection{\label{sec:level2}Atomic structures and magnetic properties}

For the $\rm VSi_2N_4$ bilayer, six highly-symmetric interlayer stacking configurations are taken into account in our calculations to search for the most stable atomic structure. The detailed information of these stacking configurations is given in Supplementary Material (SM hereafter). Given that the $\rm VSi_2N_4$ monolayer has a ferromagnetic order \cite{VSi2N4Monolayer}, both the interlayer ferromagnetic and antiferromagnetic couplings are further considered for each stacking configuration. According to the total energy calculations, it is found the $\rm AA^{\prime}$ stacking has the lowest energy among all atomic configurations. The energy of the interlayer antiferromagnetic coupling in the $\rm AA^{\prime}$ stacking is further smaller than that of the interlayer ferromagnetic one by 0.04 meV. Therefore, the antiferromagnetically coupled $\rm AA^{\prime}$ stacking is the most stable. Besides, the interlayer magnetic coupling is not strong, and it has the same order of magnitude with those of $\rm KCrS_2$ and $\rm MnBi_4Te_7$ van der Waals layers \cite{MCrS2,MnBiTe}. The interlayer magnetic order is thus likely to be switched by applied magnetic field \cite{magneticfield}, magnetic proximity effect \cite{PRB15} or other methods \cite{spin-orbit,strain-tuning}. The $\rm AA^{\prime}$ stacking with two kinds of interlayer magnetic couplings are focused on in the following calculations, and they exhibit distinct valley and spin properties. We also consider magnetic properties and valley-spin physics of $\rm VSi_2P_4$ bilayer, and its interlayer magnetic coupling is larger than that of the $\rm VSi_2N_4$ bilayer by one order of magnitude, which we will discuss in detail later. 

\begin{figure}[htbp]
\includegraphics[width=7.5cm]{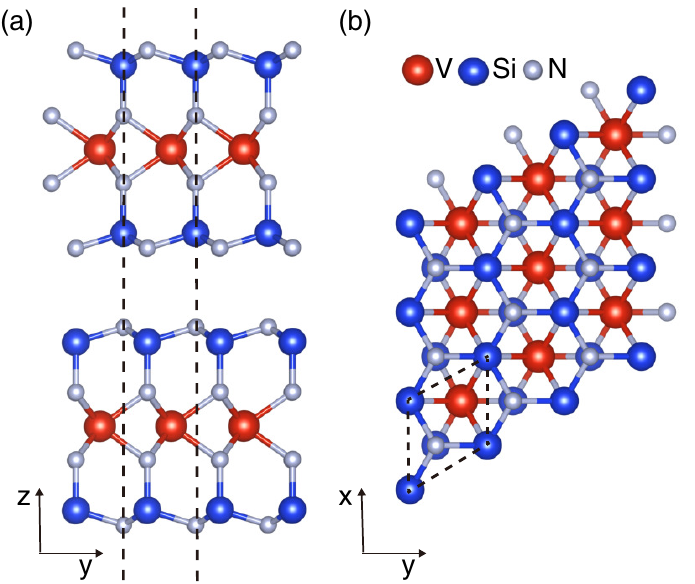}% Here is how to import EPS art
\caption{\label{fig:epsart} Atomic structure of the $\rm AA^{\prime}$-stacked $\rm VSi_2N_4$ bilayer. (a) The side view and (b) the top view. Red, blue and grey balls stand for V, Si and N atoms, respectively. The unit cell is bound by dashed lines.}
\end{figure}

Fig. 1 shows the atomic structure of the $\rm AA^{\prime}$ stacking of the $\rm VSi_2N_4$ bilayer. For each monolayer of the bilayer structure, it forms a two-dimensional hexagonal lattice, and consists of seven atomic layers stacked as N-Si-N-V-N-Si-N \cite{VSi2N4Monolayer}. 
The $\rm VN_2$ layer in the middle of the monolayer exhibits a similar structure with the 2H-$\rm MoS_2$ monolayer \cite{2012Valley}, i.e. each V atom has six neighboring N atoms forming a trigonal prism. Two SiN layers, with a buckled honeycomb lattice, are located on the top and bottom sides of the $\rm VN_2$ layer. When two $\rm VSi_2N_4$ monolayers are stacked together, the stacking configuration is determined by relative in-plane positions between the $\rm VN_2$ layers from the two monolayers. 
For the $\rm AA^{\prime}$ stacking, the V atomic layers from two $\rm VN_2$ layers coincide with each other, and the N atoms from the upper (lower) $\rm VN_2$ layer are superposed onto the centers of the hexagonal rings of the lower (upper) $\rm VN_2$ layer.

For the $\rm VSi_2N_4$ bilayer, the in-plane lattice constant of its hexagonal lattice is computed to 2.89 $\rm \AA$. The thickness of each monolayer is 6.87 $\rm \AA$, and the van der Waals gap between the two monolayers is 2.98 $\rm \AA$. 
Each V atom contributes to a magnetic moment of 1.2 $\rm \mu_B$, while magnetic moments of the other atoms are orders of magnitude smaller, which is consistent with previous results of the $\rm VSi_2N_4$ monolayer \cite{VSi2N4Monolayer}. 
The above computed structural and magnetic parameters are insensitive to the interlayer magnetic orders.

\subsection{Valley and spin properties in electronic structures}
\begin{figure}[htbp]
\includegraphics[width=8cm]{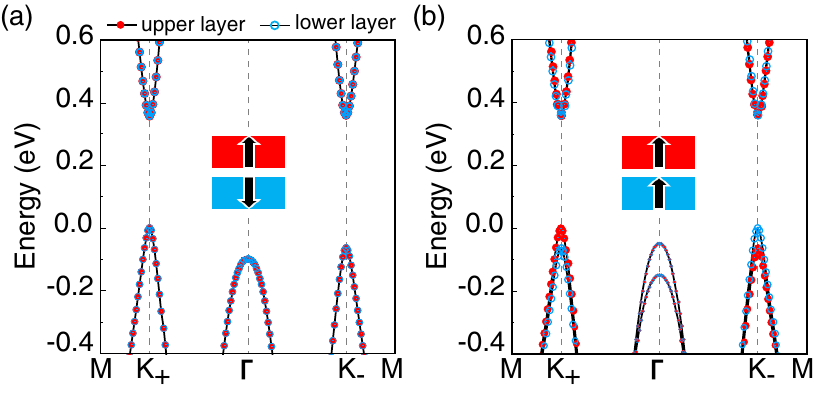}% Here is how to import EPS art
\caption{\label{fig:epsart} Electronic band structures of the $\rm AA^{\prime}$-stacked $\rm VSi_2N_4$ bilayers. (a) The interlayer antiferromagnetic coupling and (b) the ferromagnetic one. The red solid circles and blue hollow circles are used to indicate the atom-projected weights of electronic states on the V atoms from the upper and lower monolayers, resspectively. The valence band maximum is set to zero energy. The insets schematically show the interlayer magnetic order.}
\end{figure}

We then investigate the band structures and associated valley-spin properties of $\rm AA^{\prime}$-stacked $\rm VSi_2N_4$ bilayers with interlayer antiferromagnetic and ferromagnetic couplings, which are shown in Fig. 2(a) and (b) with atomic projections and the schematic depictions of the interlayer magnetic orders. In the electronic band structure of the interlayer antiferromagnetically coupled bilayer, it is seen that all bands are doubly degenerate at each crystal wave vector $\bold k $. There are two local, unequal valence band maxima (conduction band minima) at the highly symmetric $K_+$ and $K_-$ points of the hexagonal Brillouin zone. That is, a pair of inequivalent valleys appear with energy splittings for both the valence and conduction bands. According to the definition in the above section, the valley splittings have values of 64.1 meV and -1.6 meV for the valence and conduction bands, respectively, and the signs of the valley splittings can be changed by reversing the magnetization directions of both monolayers, which are similar to the case of the $\rm VSi_2N_4$ monolayer \cite{VSi2N4Monolayer}. Moreover, the direct band gaps are also unequal at $K_+$ and $K_-$ valleys, with magnitudes of 357.9 meV and 423.5 meV, respectively. Since the global valence band maximum and conduction band minimum are both located at $K_+$, the $\rm VSi_2N_4$ bilayer is a direct-band-gap semiconductor.

According to the atomic projections in Fig. 2(a), it is seen that the electronic bands at $K_{\pm}$ valleys are dominantly contributed by $ d $ orbitals of V atoms. For each pair of doubly-degenerate bands, they are plotted by red solid circles and blue hollow circles, corresponding to the contributions from the V atoms of the upper and lower monolayers, respectively. 
Further considering the spin orientation as illustrated by the spin-resolved band structures in Fig. S2 of SM, it is found that the bands composed by different monolayers exhibit opposite spins. This leads to the spin-layer locking and indicates that the doubly-degenerate bands also correspond to opposite spins. The spin degeneracy arises from the invariance of the magnetic system under the combined operations of the spatial inversion ($\mathcal{P}$) and time reversal ($\mathcal{T}$). The $\mathcal{PT}$ symmetry relates the electronic states with opposite spins at a certain $\bold k $ point, and ensures that they have the same energy \cite{13PANS}. 

In contrast to the valley splittings in the interlayer antiferromagnetic coupling, there are two degenerate valleys at $K_{\pm}$ points for the $\rm AA^{\prime}$-stacked bilayer with the interlayer ferromagnetic coupling. Owing to the valley degeneracy, the direct band gaps at $K_{\pm}$ are equal with a magnitude of 357.7 meV, and the bilayer is still a direct-band-gap semiconductor. Moreover, the electronic bands are no longer degenerate. Two valence bands and two conduction bands appear at each valley within the energy range of Fig. 2(b). The splitting between the first valence (conduction) band near the Fermi level and the second one is 64.1 meV (1.6 meV).

The atomic projections in Fig. 2(b) further demonstrate that for the $K_+$ ($K_-$) valley, the first (second) valence and conduction bands with red solid circles are mainly contributed by the V atom from the upper monolayer, while the V atom from the lower monolayer dominantly contributes to the second (first) valence and conduction bands with blue hollow circles. Therefore, for each band, the electronic states at $K_+$ and $K_-$ valleys have distinct distributions at two monolayers, indicating a characteristic of the valley-layer coupling. When concentrating on the red fat bands contributed by the upper monolayer, valley splittings of 64.1 meV and -1.6 meV are found for the valence and conduction bands, respectively, which agrees with the splittings in single $\rm VSi_2N_4$ monolayer \cite{VSi2N4Monolayer}. For the blue fat bands contributed by the lower monolayer, the valley splittings have the same magnitude but opposite signs. Moreover, it is found that the bands composed by different monolayers have the same spin for the interlayer ferromagnetic coupling, by the spin projection in Fig. S2 of SM. The degenerate valleys with the single spin are ensured by the spatial inversion symmetry of the bilayer system, which relates the electronic states with the same spin but at opposite crystal wave vectors.

The $\rm AA^{\prime}$-stacked $\rm VSi_2N_4$ bilayers with the interlayer antiferromagnetic and ferromagnetic couplings exhibit distinct valley degeneracy and spin degeneracy. Besides, there are the spin-layer coupling in the antiferromagnetic order and the valley-layer coupling in the ferromagnetic order. Therefore, the selection of the layer is likely to lead to valley or spin polarization. In the followings, we will investigate the roles of applied vertical electric field in tuning electronic band structure of the $\rm VSi_2N_4$ bilayers.

\subsection{The roles of applied electric field}

We first study the effects of the electric field on the antiferromagnetically coupled bilayer with the valley splitting and spin degeneracy. When adding a positive electric field that points upwards, it is found that each pair of degenerate bands in Fig. 2(a) is splitted into two non-degenerate bands with  atomic contributions from different monolayers and opposite spins, as shown in Fig. 3(a). Specifically speaking, the first (second) valence band generated by the splitting is contributed by only the upper (lower) monolayer with up (down) spin at both valleys, and the first (second) conduction band has the same atomic contribution and spin orientation as the second (first) valence band. The first valence and conduction bands thus have opposite spins at each valley, indicating that the bilayer is a bipolar magnetic semiconductor \cite{BMS}.
The spin splittings above arise from the up shift of the spin-up red bands in Fig. 3(a) with respect to the spin-down blue bands, which is also demonstrated schematically in Fig. 3(b). The relative band shift is due to the electric potential difference between the upper and lower monolayers induced by the applied electric field and the spin-layer coupling. From the view of the symmetry, it is the electric field that breaks the $\mathcal{PT}$ symmetry and associated spin degeneracy.

\begin{figure}[htbp]
\includegraphics[width=9cm]{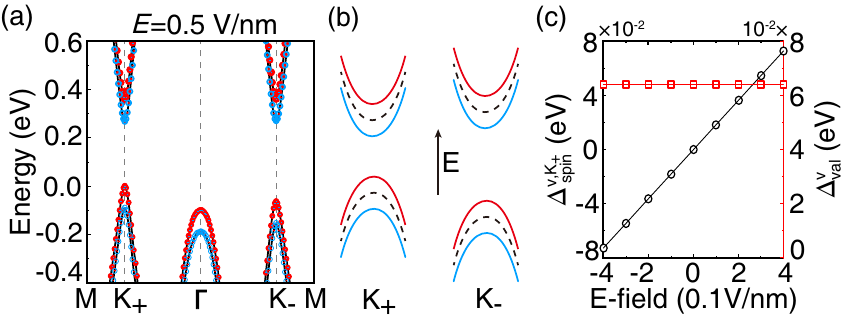}% Here is how to import EPS art
\caption{\label{fig:epsart} Electronic properties of the interlayer antiferromagnetically coupled bilayer, under applied electric field. (a) The calculated atom-projected band structure with an electric field of 0.5 V/nm. (b) Schematic depiction of the band shift. The black dashed lines and red/blue solid lines represent the degenerate bands 
 and the splitted bands at $K_\pm$ valleys, before and after adding the electric field, respectively. The upward arrow indicates the direction of the electric field. (c) The valley and spin splittings as functions of the electric field. While the valley splitting is that of the valence band, the spin splitting corresponds to the valence band at $K_+$ valley.}
\end{figure}
\vspace{0.3cm}
Moreover, the evolutions of the spin splitting and the valley   splitting with the electric field are demonstrated in Fig. 3(c). As the strength of the electric field is enhanced, the spin splittings are linearly increased for both valence and conduction bands at two valleys, by the same rate of about 18.2 meV per 0.1 V/nm. Correspondingly, the direct band gap at two valleys linearly decreases with the electric field by the above rate. In contrast, the valley splittings of the bilayer do not change with the electric field, since the orbital component of the valence/conduction bands at two valleys come from the same monolayer and the valley splitting of the single monolayer is insensitive to the applied electric field. Therefore, the presence of the applied electric field leads to the spin splitting and has no influence on the valley splittings. Besides, the sign of the spin splittings is determined by the direction of the electric field. When applying a negative electric field that points downwards, the induced spin splittings are reversed compared with the case of the positive electric field.  

\begin{figure}[htbp]
\includegraphics[width=9cm]{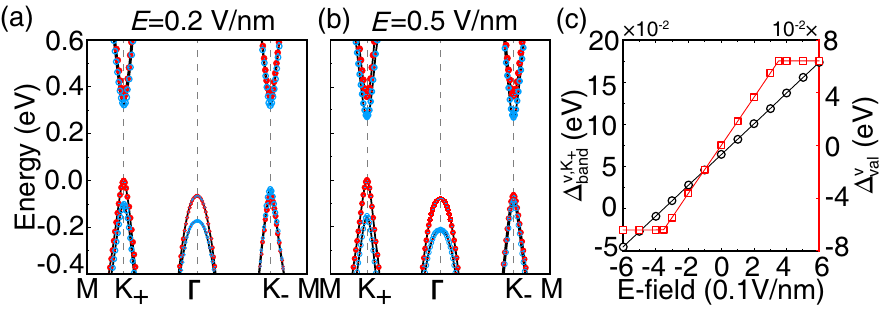}% Here is how to import EPS art
\caption{\label{fig:epsart}Electronic properties of the interlayer ferromagnetically coupled bilayer, under applied electric field. (a) and (b) The atom-projected band structures with electric fields of 0.2 V/nm and 0.5 V/nm, respectively. (c) The band and valley splittings as functions of the electric field. The band splitting, $\Delta_{\text {band }}^{v, K_+}$, corresponds to the relative shift between the first and second valence bands at $K_+$ valley, while the valley splitting is that of the valence band.}
\end{figure}
On the other hand, the electric field is introduced into the ferromagnetically coupled bilayer with the valley degeneracy. Under the action of the positive electric field, the red bands contributed by the upper monolayer move upwards with respect to the blue bands contributed by the lower monolayer, which is similar to the above case and shown in Figs. 4(a) and (b) with two representative electric fields. Since the degenerate bands from two valleys in Fig. 2(b) correspond to different monolayers of the pristine bilayer, i.e. the valley-layer coupling, the band shifts under the electric field results in the valley splitting.

The size of the valley splitting depends on the strength of the electric field. We first consider the valley splitting of the valence band, as shown in Fig. 4(c). When the positive electric field is within the range of 0-0.35 V/nm, the splitting is enhanced linearly with the electric field, up to 64.1 meV. When the electric field continues increasing, the splitting is unchanged. The electronic bands in Figs. 4(a) and (b) correspond to the two strength ranges of the electric field, respectively. The trend change is due to a critical electric field of 0.35 V/nm that gives rise to a band exchange between the first and second valence bands at $K_-$ valley. Beyond the critical electric field, the first valence bands at two valleys are both contributed by the same monolayer. As a result, the valley-layer coupling is broken and the valley splitting of the valence band is no longer varied with the electric field. The calculated critical electric field can be well estimated by dividing the aforementioned band splitting between the first and second valence bands of the pristine bilayer by the relative shift rate of the bands from two monolayers under the electric field. Given that the band splitting above is 64.1 meV and the band shift rate is 18.2 meV per 0.1 V/nm, the estimated critical electric field is also 0.35 V/nm. Moreover, the reversal of the electric field changes the sign of the valley splitting.

Further considering the conduction band, its valley splitting has the similar trend as that of the valence band. However, the critical electric field for the conduction band is much smaller, with a value of 9 $\times$ $10^{-3}$ V/nm. It is because the splitting between the first and second conduction bands is much smaller, compared with that of the valence bands. Besides, owing to the valley splitting induced by the electric field, there is a transition from the direct-band-gap semiconductor to the indirect-band-gap one. For the positive (negative) electric field, the valence band maximum and conduction band minimum are located at $K_+$ ($K_-$) and $K_-$ ($K_+$) valleys, respectively.  
\section{Discussions}

Distinct band structures and subsequent electric control in $\rm AA^{\prime}$-stacked bilayers lead to various Hall effects. For the interlayer antiferromagnetic coupling, a moderate carrier doping will occupy one valley due to the presence of the valley splitting in Fig. 2(a). The spin degeneracy ensured by the $\mathcal{PT}$ symmetry gives rise to opposite anomalous Hall currents generated by the carriers with opposite spins from the occupied valley, owing to their opposite Berry curvatures. That is, a spin Hall effect will appear when the vertical electric field is absent, similar to the case of honeycomb lattices with N{\'e}el antiferromagnetism \cite{13PANS}. 
For the interlayer ferromagnetic coupling, the valley degeneracy in Fig. 2(b) results in opposite anomalous Hall currents from different valleys but with the same spin by carriers doping, indicating a valley Hall effect. 
Therefore, the bilayer with the interlayer ferromagnetic order can be regards as single-spin version of the two-valley semiconductor. On the other hand, after applying a vertical electric field, neither the valley degeneracy nor spin degeneracy is present in the antiferromagnetic/ferromagnetic coupled bilayer, which leads to a single anomalous Hall current with certain indices of valley and spin. Moreover, the indices can be selectively switched by applying an opposite electric field or reversing the magnetization direction.

\begin{figure}[htbp]
\includegraphics[width=9cm]{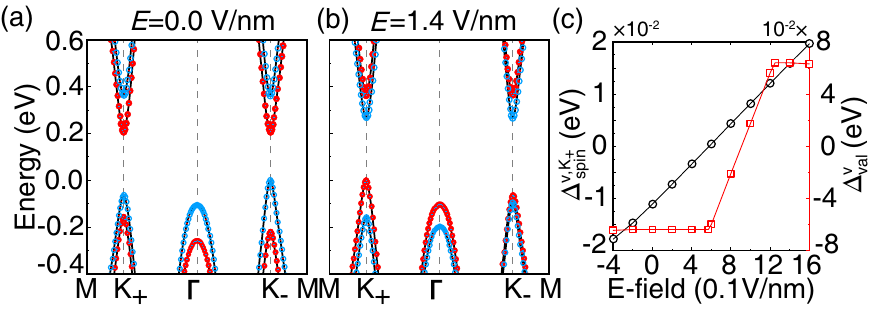}% Here is how to import EPS art
\caption{\label{fig:epsart}Electronic properties of the AB-stacked bilayer with the interlayer antiferromagnetic order. (a) and (b) The atom-projected band structures without any electric field and with an electric field of 1.4 V/nm, respectively. (c) The valley and spin splittings as functions of the electric field. While the valley splitting is that of the valence band, the spin splitting corresponds to the valence band at $K_+$ valley.}
\end{figure}
\setlength{\belowcaptionskip}{2cm} 
%\setlength{\abovecaptionskip}{0.cm}
%\vspace{2cm}

Besides the aforementioned $\rm AA^{\prime}$  stacking, the other stable stackings also exhibit different valley and spin properties. For the second stable configuration-the AB stacking, its interlayer antiferromagnetic coupling gives a band structure with both spin splittings and valley splittings, as shown in Fig. 5(a). Under the action of applied electric field, the valley and spin splittings are reversed in Fig. 5(b). Therefore, spin-polarized and valley-polarized anomalous Hall currents can appear in the configuration, and the polarization can be changed by the electric field. Fig. 5(c) further demonstrates the evolutions of the splittings as functions of the electric field. When applying an electric field in the range of 0.57-1.25 V/nm, the valley splitting of the valence band linearly increases from -64.1 meV to +64.1 meV. With a positive electric field of smaller than 0.57 V/nm (more than 1.25 V/nm), the valley splitting is insensitive to the electric field, with a value of -64.1 meV (64.1 meV). The negative electric field also has no influence on the valley splitting. Therefore, the sign of the valley splitting can be reversed by a unidirectional electric field, which is the third kind of the change trend of the valley splitting, in contrast to the cases of the $\rm AA^{\prime}$-stacked bilayer with two kinds of the interlayer magnetic couplings in Figs. 3(c) and 4(c). Besides, the valley splitting of the conduction band exhibits the similar trend, but with much smaller electric field. Different from the valley splittings, the spin splittings linearly changes within the strength range of the electric field considered here. Moreover, the interlayer antiferromagnetically coupled AA stacking has a similar band structure as Fig. 2(b), except that its bands contributed by different monolayers also correspond to different spins. Besides the valley-layer coupling in Fig. 2(b), there are also spin-layer coupling and valley-spin coupling for the AA stacking. More details on different stackings are summarized in SM.

The above intriguing physics in $\rm VSi_2N_4$ bilayers have generalizations to other magnetic van der Waals bilayers, such as the bilayers of $\rm VSi_2P_4$ \cite{VSi2P4}, $\rm CrSi_2N_4$ \cite{CrSi2N4}, $\rm CrSi_2P_4$ \cite{CrSi2N4}, 2H-$\rm VS_2$ \cite{VS2bilayer},  2H-$\rm VSe_2$ \cite{2VSe2} and so on. Taking the $\rm VSi_2P_4$ bilayer for another example, we also computed its atomic and electronic properties. It is found that the $\rm VSi_2P_4$ bilayer has similar characteristics with the $\rm VSi_2N_4$ bilayer, including the most stable configuration, valley and spin degeneracies, and responses to applied electric field, which are given in SM. Compared with $\rm VSi_2N_4$, the $\rm VSi_2P_4$ bilayer has much larger interlayer magnetic coupling, according to an energy difference of 0.4 meV between the antiferromagnetically and ferromagnetically coupled bilayers. The band shift of $\rm VSi_2P_4$ bilayer is also approximately linear in the strength of the electric field, but with a smaller rate that is about a half of the one of the $\rm VSi_2N_4$ bilayer.

\section{Conclusion}

In summary, by first-principles calculations we have demonstrated rich valley-spin physics in $\rm VSi_2N_4$ bilayers and subsequent electric controls. Considering the most stable $\rm AA^{\prime}$  stacking, there are the spin degeneracy and the valley splitting in the electronic band structure of the interlayer antiferromagnetic coupling, while the valley degeneracy and the band splitting appear for the interlayer ferromagnetic coupling. As a result, two kinds of the interlayer magnetic orders give rise to spin and valley Hall effects, respectively. Besides, according to the atomic projections, they exhibit the spin-layer coupling and the valley-layer coupling, respectively, enabling the electric controls of spin and valley by selectively expressing the layer degree of freedom. Under the action of applied vertical electric field, both the spin degeneracy in the antiferromagnetic order and the valley degeneracy in the ferromagnetic order are lifted, where corresponding splittings are highly tunable for both the magnitude and the sign. The resulted single anomalous Hall current can selectively arise from any valley with any spin. Moreover, the other stacking orders provide more choices of valley and spin degeneracies, and more material realizations can be expected among a number of magnetic bilayers with inequivalent valleys. This findings give full play to the valley and spin degrees of freedom with the help of layered structure and associated electric control, and offers various possibilities for advanced spintronic and valleytronic devices.

\begin{acknowledgments}
We acknowledge ﬁnancial supports from the National Natural Science Foundation of China Grant 11904173 and the Jiangsu Specially-Appointed Professor Program.
\end{acknowledgments}

\end{document}